\documentclass{amsproc}
\usepackage{graphicx}
\usepackage{amssymb,amsmath,latexsym}

\newtheorem{theorem}{Theorem}[section]
\newtheorem{lemma}[theorem]{Lemma}

\theoremstyle{definition}
\newtheorem{definition}[theorem]{Definition}
\newtheorem{example}[theorem]{Example}
\newtheorem{xca}[theorem]{Exercise}

\theoremstyle{remark}
\newtheorem{remark}[theorem]{Remark}

\numberwithin{equation}{section}

\newcommand{\abs}[1]{\lvert#1\rvert}

\newcommand{\bea}{\begin{eqnarray}}
\newcommand{\ena}{\end{eqnarray}}

\newcommand{\beano}{\begin{eqnarray*}}
\newcommand{\enano}{\end{eqnarray*}}

\newcommand{\bee}{\begin{enumerate}}
\newcommand{\ene}{\end{enumerate}}

\newcommand{\bei}{\begin{itemize}}
\newcommand{\eni}{\end{itemize}}

\newcommand{\ket}[1]{\mid #1 >}
\newcommand{\bra}[1]{<#1 \mid}
\newcommand{\matrixdd}[4]{\left(\begin{array}{cc}#1&#2\\#3&#4\\ 
\end{array} \right)}

\def\R{\mathbb{R}}

\def\C{\mathbb{C}}

\def\I{\mathbb{I}}

\newcommand{\blankbox}[2]{%
  \parbox{\columnwidth}{\centering
    \setlength{\fboxsep}{0pt}%
    \fbox{\raisebox{0pt}[#2]{\hspace{#1}}}%
  }%
}

\begin{document}

\title{ Examples of Berezin-Toeplitz Quantization: Finite sets and  Unit Interval }

\author{J.-P. Gazeau}
\address{ LPTMC and F\'ed\'eration 
de Recherches
Astroparticules et Cosmologie,
Boite 7020, Universit\'e Paris 7
Denis Diderot,
F-75251 Paris Cedex 05, France}
\email{gazeau@ccr.jussieu.fr }

\author{T. Garidi}
\address{ LPTMC and F\'ed\'eration 
de Recherches
Astroparticules et Cosmologie,
Boite 7020, Universit\'e Paris 7
Denis Diderot,
F-75251 Paris Cedex 05, France}
\email{garidi@ccr.jussieu.fr}

\author{E. Huguet}
\address{ D\'epartement d'Astrophysique Stellaire et Galactique and F\'ed\'eration 
de Recherches
Astroparticules et Cosmologie, 
Observatoire de Paris-Meudon, 92195 Meudon, France}
\email{eric@bluesette.obspm.fr}

\author{M. Lachi\`eze Rey}
\address{ Service d'Astrophysique du CEA et F\'ed\'eration 
de Recherches
Astroparticules et Cosmologie, CEA Saclay,
91191 Gif sur Yvette Cedex, France  }
\email{marclr@cea.fr}

\author{J. Renaud}
\address{LPTMC and F\'ed\'eration 
de Recherches
Astroparticules et Cosmologie, Boite 7020, Universit\'e Paris 7
Denis Diderot,
F-75251 Paris Cedex 05, France}
\email{renaud@ccr.jussieu.fr }

\date{today}
\dedicatory{In memory of Bob Sharp}


\begin{abstract}
We present a quantization scheme of an arbitrary measure space based
on overcomplete families of states and generalizing the Klauder
and the Berezin-Toeplitz approaches. This scheme could reveal itself
as an efficient tool for quantizing physical systems for which 
more traditional methods like geometric quantization are uneasy to implement.
The procedure is illustrated by (mostly two-dimensional)
elementary examples in which the measure space is a $N$-element set and the unit interval. 
Spaces of states for the $N$-element set and the unit interval are the
2-dimensional euclidean $\R^2$ and hermitian $\C^2$ planes.
\end{abstract}

\maketitle


.

\section{Quantum processing of a measure space}
Quantum Physics and Signal Analysis have many aspects in common. As
a departure point of their respective formalism, one finds a {\it
raw} set $X=\{x\}$ of basic parameters or data. This set may be a classical
phase space in the former case whereas it might be a temporal line or
a time-frequency half-plane in the latter one. In reality it can be any
set of data accessible to observation. The minimal significant
structure one requires of it is the existence of a measure
$\mu(dx)$, together with a $\sigma$-algebra of measurable subsets. As
a measure space, $X$ will be given the name of an {\it observation}
set in the present context, and the existence of a measure provides
us with a statistical reading of the set of measurable real or
complex valued functions $f(x)$ on $X$: computing for instance
average values on subsets with bounded measure. Actually, both
approaches deal with quadratic mean values and correlation/convolution
involving signal pairs, and the natural frameworks of studies are the
real (Euclidean) or complex (Hilbert) spaces, $L^2(X, \mu) \equiv
L_{\R}^2(X, \mu)$ or $L_{\C}^2(X, \mu)$ of square integrable
functions $f(x)$ on the observation set $X$: $\int_X \vert
f(x)\vert^2 \, \mu(dx) < \infty $. One will speak of {\it
finite-energy} signal in Signal Analysis and of quantum state in
Quantum Mechanics. However, it is precisely at this stage that
``quantum processing'' of $X$ differs from signal processing on at
least three points : \bee
\item not all square integrable functions are eligible as quantum
states, 
\item a quantum state is defined up to a nonzero factor,
\item those ones among functions $f(x)$ that are eligible as quantum
states with  unit norm, $\int_X \vert f(x)\vert^2\, \mu(dx)
= 1$, give rise to a probability interpretation: $X \supset \Delta
\rightarrow \int_{\Delta} \vert f(x) \vert^2 \mu(dx)$ is a
probability measure interpretable in terms of localisation in the
measurable $\Delta$. This is  inherent to the computing of mean values of
quantum observables, (essentially) self-adjoint operators with domain
included in the set of quantum states. \ene
The first point lies at the heart of the {\it quantization} problem :
what is the more or less canonical procedure allowing to select
quantum states among simple signals? In other words, how to select
the right (projective) Hilbert space  ${\mathcal H}$, a closed
subspace of $L^2(X, \mu)$, or equivalently the corresponding
orthogonal projecteur $\I_{{\mathcal H}}$?

In various cicumstances, this question may be  answered through the
selection, among elements of $L^2(X, \mu)$, of an orthonormal set
$\mathcal{S}_N = \{ \phi_n(x) \}_{n = 1}^N$, $N$ being finite or
infinite, which spans, by definition,  the separable Hilbert subspace
${\mathcal H}
\equiv {\mathcal H}_N$. Furthermore, and this is a crucial assumption \cite{aag,klau2,csbook}, we require that
\begin{equation} 
{\mathcal N} (x) \equiv \sum_n \vert \phi_n (x) \vert^2 <
\infty \ \mbox{almost everywhere}. \label{factor}
\end{equation}
Of course, if $N$ is finite the above condition is trivially checked.

We then consider the   family of states $\{ | x \rangle \}_{x\in X}$ through the
following linear superpositions: \begin{equation}
| x\rangle \equiv \frac{1}{\sqrt{{\mathcal N} (x)}} \sum_n \phi_n (x)
| n\rangle, \label{cs}
\end{equation}
in which the ket $| n\rangle$ designates the element $\phi_n(x)$ in a
``Fock'' notation.
This defines an injective map
\begin{equation}
  X \ni x
\rightarrow | x \rangle \in {\mathcal H}_N    \in {\mathcal H}_N 
\end{equation}(in Dirac notations),
and it is not difficult to check that states
(\ref{cs}) are \textit{coherent} in the sense that they obey
the following two conditions:
\bei
\item {\bf Normalisation}
\begin{equation}\langle \, x\, | x \rangle = 1, \label{norma}
\end{equation}\item {\bf Resolution of the unity in ${\mathcal H}_N$}
\begin{equation}\int_X | x\rangle \langle x  | \,\, \nu(dx)=
\I_{{\mathcal H}_N}, \label{iden}
\end{equation}where $\nu(dx) = {\mathcal N}(x)\,\mu(dx)$ is another
measure on $X$, absolutely continuous with respect to $\mu(dx)$. The coherent
states (\ref{cs}) form in general an overcomplete (continuous) basis of ${\mathcal H}$.
\eni
The resolution of the unity in ${\mathcal H}_N$ can alternatively
been understood in terms of the scalar product $\langle \, x\, | x'
\rangle$ of two states of the family. Indeed, (\ref{iden}) implies
that, to any vector $| \phi \rangle$ in ${\mathcal H}_N$ one can
(anti-)isometrically associate the function \begin{equation}
\phi^{\star}(x) \equiv
\sqrt{\mathcal N(x)}\langle x\, | \phi \rangle
\end{equation} in $L^2(X, \mu)$, and
this function obeys \begin{equation}
\phi^{\star}(x) = \int_X \sqrt{\mathcal
N(x)\mathcal N(x')} \langle x| x' \rangle \phi^{\star}(x')\, \mu(dx') .
\end{equation} Hence, ${\mathcal H_N}$ is (anti-) isometric to a reproducing
Hilbert space with kernel \begin{equation}
{\mathcal K}(x,x') = \sqrt{\mathcal N(x)\mathcal N(x')}
\langle x\, | x' \rangle,
\end{equation} and the latter assumes finite diagonal
values ({\it a.e.}), ${\mathcal K}(x,x) = \mathcal N(x)$, by construction.

A {\it classical} observable is  a
function $f(x)$ on $X$ having specific properties in relationship
with some supplementary structure allocated to $X$, topology,
geometry or something else. Its  quantization \cite{klau1,ber}   simply consists in 
associating to $f(x)$
the operator
\begin{equation}A_f := \int_X f(x) | x\rangle \langle x| \, \nu(dx).
\label{oper}
\end{equation}In this context, $f(x)$ is said upper (or
contravariant) symbol of the operator $A_f$ and denoted by $f = \hat{A}_f $, whereas the mean value
$\langle x| f(x) | x\rangle$ is said lower (or covariant) symbol of
$A_f$ \cite{ber,csfks} and denoted by $\check{A}_f$.
Through this approach, one can say
that a quantization of the observation set is
in one-to-one correspondence with the choice of a frame in the sense
of (\ref{norma}) and (\ref{iden}). To a certain extent, a
quantization scheme consists in adopting a certain point of view in
dealing with $X$. This frame can be discrete, continuous, depending
on the topology furthermore allocated to the set $X$, and it can be
overcomplete, of course. The validity of a precise frame choice is
asserted by comparing spectral characteristics of quantum observables
$A_f$ with  data issued from a predefined experimental protocole. Of course, operators
acting in  ${\mathcal H}_N$ are not all of them  of the ``diagonal'' type $A_f$, and many different
classical $f(x)$'s can give rise to the \textit{same} operator $A_f$. The frame  should be 
complete or rich enough in order to meet all experimental possibilities determined by the protocole. 

Let us illustrate the above construction with the well-known Klauder-Glauber-Sudarshan coherent
states \cite{csfks} and the subsequent so-called canonical quantization. 
The observation set $X$ is the classical phase space $\R^2 \simeq \C = \{ x
\equiv z = 
\frac{1}{\sqrt2}(q+ip) \}$ (in complex notations) of a particle with one degree of freedom. The
measure on $X$ is Gaussian, $\mu(dx) = \frac{1}{\pi}\, e^{-\vert z \vert^2}\, d^2 z $ where $d^2
z$ is the Lebesgue measure of the plane. The functions $\phi_n (x)$ are the normalised
powers of the complex variable $z$, $\phi_n (x) \equiv \frac{z^n}{\sqrt{n!}}$, so that the
Hilbert subspace ${\mathcal H}$ is the so-called Fock-Bargmann space of all entire functions that are
square integrable with respect to the Gaussian measure.  Since $
 \sum_n \frac{\vert z \vert^2}{n!} =  e^{\vert z \vert^2}$, the coherent states read
\begin{equation}| z\rangle = e^{-\frac{\vert z \vert^2}{2}} \sum_n  \frac{z^n}{\sqrt{n!}}| n\rangle,
\label{scs}
\end{equation}and one easily checks the normalisation and unity resolution:
\begin{equation}\langle z\, | z \rangle = 1,\  \ 
\frac{1}{\pi}\int_{\C}  | z\rangle \langle z| \, d^2 z= \I_{{\mathcal H}},
\label{pscs}
\end{equation}Note that the reproducing kernel is simply given by $e^{\bar{z}z'}$.
The quantization of the observation set is hence achieved by selecting in the original Hilbert
space $L^2(\C, \frac{1}{\pi}e^{-\vert z \vert^2}\, d^2 z)$ all holomorphic entire functions, which geometric
quantization specialists would call a choice of polarization.  Quantum operators acting on ${\mathcal
H}$ are yielded by using (\ref{oper}). We thus have for the most basic one,
\begin{equation}\frac{1}{\pi}\int_{\C}  z\, | z\rangle \langle z| \, d^2 z = \sum_n \sqrt{n+1} 
| n\rangle \langle n+1| \equiv a,
\label{low}
\end{equation}which is the lowering operator, $a | n\rangle = \sqrt{n} | n - 1\rangle$. Its adjoint
$a^{\dagger}$ is obtained by replacing $z$ by $\bar{z}$ in (\ref{low}), and we get the
factorisation $N = a^{\dagger}a$ for the number operator, together with the  commutation rule $\lbrack a, a^{\dagger}
\rbrack = \I_{{\mathcal H}}$. Also note that $ a^{\dagger}$ and $a$ realize on ${\mathcal H}$ as
multiplication operator and derivation operator respectively, $a^{\dagger}f(z) = zf(z), \ af(z)
= df(z)/dz$. From $q = \frac{1}{2}(z +
\bar{z})$ et  $p = \frac{1}{2i}(z - \bar{z})$, one easily infers by linearity that $q$ and $p$
are   upper symbols for $\frac{1}{2}(a + a^{\dagger}) \equiv Q$ and $\frac{1}{2i}(a - a^{\dagger})
\equiv P$ respectively. In consequence,  the self-adjoint operators $Q$ and $P$ obey the canonical
commutation rule $\lbrack Q, P \rbrack = i\I_{{\mathcal H}}$, and for this reason fully deserve  the
name of position and momentum operators of the usual (galilean) quantum mechanics, together with
all  localisation properties specific to the latter.

The next examples which are presented in this paper are, although elementary, rather 
 unusual. In particular, we start with observation sets which are not necessarily  phase spaces, and
such sets are far from having  any physical meaning in the common sense. 
We first consider a two-dimensional quantization of a $N$-element set 
which leads, for $N\geq 4$,  to a Pauli  algebra of observables. We then study two-dimensional (and higher-dimensional)
quantizations of the unit segment. In the conclusion, we shall mention some questions of physical interest  which  are currently 
under investigation.

\section{Quantum processing of a $N$-element set}
An elementary (but not trivial!) exercise for illustrating  the quantization scheme
introduced in the previous section involves an arbitrary $N$-element set $X= \{x_i \}$ 
as observation set. An arbitrary non-degenerate measure on it is given by a sum of  Dirac measures:
\begin{equation}
\mu(dx) = \sum_{i=1}^N a_i \delta_{\{x_i\}}, \ a_i > 0.
\end{equation}
The Hilbert space $L^2(X,\mu)$ is simply isomorphic to $\C^N$. 
An obvious  orthonormal basis is given by $\left\{\frac{1}{\sqrt{a_i}} \chi_{\{x_i\}}(x), \ i=1, \cdots , N\right\}$, where
$\chi_{\{a\}}$ is the characteristic function of the singleton $\{a\}$. 
We now consider the two-element orthonormal set $\{ \phi_1\equiv \phi_{\alpha} \equiv| {\pmb \alpha }\rangle ,
\phi_2 \equiv \phi_{\beta} \equiv | {\pmb \beta } \rangle\} $ defined in the most generic way by:
\begin{equation} \label{osN}
\phi_{\alpha}(x)  =  \sum_{i=1}^N \alpha_i \frac{1}{\sqrt{a_i}} \chi_{\{x_i\}}(x), 
\ \
\phi_{\beta}(x)  = \sum_{i=1}^N \beta_i \frac{1}{\sqrt{a_i}} \chi_{\{x_i\}}(x),
\end{equation}
where complex coefficients  $\alpha_i$ and $\beta_i$ obey
\begin{equation}
\sum_{i=1}^N \vert \alpha_i \vert^2 = 1 = \sum_{i=1}^N \vert \beta_i \vert^2, \  \sum_{i=1}^N  \alpha_i \overline{\beta_i} = 0.
\end{equation}
In a Hermitian geometry language, our choice of  $\{\phi_{\alpha}(,
\phi_{\beta}( \} $ amounts to selecting in $\C^N$ the two  orthonormal
vectors 
${\pmb \alpha }= \{\alpha_i \}, {\pmb \beta }= \{\beta_i\}$, and this justifies our notations for indices.

It follows the expression for the coherent states:
\begin{equation}\label{Ncs}
| x \rangle = \frac{1}{\sqrt{\mathcal N(x)}} \left\lbrack \phi_{\alpha}(x) ~| {\pmb \alpha }\rangle +
\phi_{\beta}(x) ~| {\pmb \beta} \rangle \right\rbrack,
\end{equation}
in which $\mathcal N(x)$ is given by
\begin{equation}
\mathcal N(x) = \sum_{i=1}^N \frac{\vert \alpha_i \vert^2 + \vert \beta_i \vert^2}{a_i}  \chi_{\{x_i\}}(x).
\end{equation}
The resolution of unity (\ref{ident}) here reads as:
\begin{equation}
\label{ }
\I = \sum_{i=1}^N \left( \vert \alpha_j \vert^2 + \vert \beta_j \vert^2 \right) | x_i \rangle \langle x_i |
\end{equation}
The overlap between two coherent states is given by the following kernel :
\begin{equation}
\label{ overlap}
\langle x_i | x_j \rangle = \frac{\overline{\alpha_i}\alpha_j + \overline{\beta_i}\beta_j }{\sqrt{\vert \alpha_i \vert^2 + \vert \beta_i \vert^2}\sqrt{\vert \alpha_j \vert^2 + \vert \beta_j \vert^2}}.
\end{equation}
To any real-valued function $f(x)$ on $X$, {\it i.e.} to any vector ${\pmb f} \equiv (f(x_i))$ in $\R^N$, there corresponds
the following  hermitian operator $A_f$ in $\C^2$, expressed in matrix form with respect to the orthonormal basis
(\ref{osN}):
\begin{equation}\label{aefN}
\begin{split}
A_f &= \int_X \mu(dx)\,\mathcal N(x)\, f(x) | x \rangle \langle x | \\ & =
\begin{pmatrix}
\sum_{i=1}^N \vert \alpha_i \vert^2 f(x_i) & \sum_{i=1}^N  \alpha_i \overline{\beta_i} f(x_i) \\ 
\sum_{i=1}^N  \overline{\alpha_i} \beta_i f(x_i) 
& \sum_{i=1}^N \vert \beta_i \vert^2 f(x_i) \end{pmatrix} \equiv  \begin{pmatrix}
\langle  {\pmb F} \rangle_{\pmb \alpha } & \langle{\pmb \beta } |  {\pmb F} | {\pmb \alpha }\rangle \\ 
\langle{\pmb \alpha } |  {\pmb F} | {\pmb \beta }\rangle & \langle  {\pmb F} \rangle_{\pmb \beta } \end{pmatrix},
\end{split}
\end{equation}
where ${\pmb F}$ holds for the diagonal matrix $\left\{ (f(x_i))\right\}$.
It is clear that, for a generic choice of the complex $\alpha_i$'s and $\beta_i$'s, 
all possible hermitian $2\times2$-matrices can
be obtained in this way if $N\geq 4$. By {\it generic} we mean that the following $4\times N$-real
matrix
\begin{equation}\label{matc}
\mathcal C = \begin{pmatrix}
 \vert \alpha_1 \vert^2 & \vert \alpha_2 \vert^2 & \cdots & \vert \alpha_N \vert^2  \\
\vert \beta_1 \vert^2 & \vert \beta_2 \vert^2 & \cdots & \vert \beta_N \vert^2 \\ 
 \Re( \alpha_1 \overline{\beta_1}) &  \Re( \alpha_2 \overline{\beta_2}) & \cdots &  \Re( \alpha_N \overline{\beta_N}) \\ 
  \Im( \alpha_1 \overline{\beta_1}) &  \Im( \alpha_2 \overline{\beta_2}) & \cdots &  \Im( \alpha_N \overline{\beta_N}) 
\end{pmatrix}
\end{equation}
 has rank equal to 4. The case $N=4$ with $\det \mathcal C \not= 0 $ is particularly interesting since then one has uniqueness of
upper symbols of Pauli matrices  $\sigma_{1}   = \bigl(
\begin{smallmatrix} 0 & 1 \\ 1 & 0 \end{smallmatrix} \bigr) ,\  \sigma_{2} = \bigl(
\begin{smallmatrix} 0 & -i \\ i & 0 \end{smallmatrix} \bigr) ,\  \sigma_{3} = \bigl(
\begin{smallmatrix} 1 & 0 \\ 0 & -1 \end{smallmatrix} \bigr) ,\ \sigma_{0} =Ê\I$, which form a basis of the four-dimensional Lie
algebra of
 complex Hermitian $2\times 2$-matrices. As a matter of fact, the operator (\ref{aefN}) decomposes with respect to this basis
as:
\begin{equation}
A_f = \langle f \rangle_+ \sigma_0 + \langle f \rangle_- \sigma_3 + \Re\left(\langle{\pmb \beta } |  {\pmb F} | {\pmb \alpha }\rangle \right)\sigma_1 - \Im\left(\langle{\pmb \beta } |  {\pmb F} | {\pmb \alpha }\rangle\right)\sigma_2,
\end{equation}
where the symbols $\langle f \rangle$ stand for the following averagings:
\begin{equation}\label{avf}
\langle f \rangle_{\pm} = \frac{1}{2}\,\sum_{i=1}^N \left(\vert \alpha_i \vert^2 \pm \vert \beta_i \vert^2\right)f(x_i) =  \frac{1}{2}\left( \langle  {\pmb F} \rangle_{\pmb \alpha }
\pm \langle  {\pmb F} \rangle_{\pmb \beta } \right).
\end{equation}
Note that $\langle f \rangle_{+}$ alone has a meanvalue status, precisely with respect to the probability distribution 
\begin{equation}\label{probf}
p_i = \frac{1}{2}\,\left(\vert \alpha_i \vert^2 + \vert \beta_i \vert^2\right).
\end{equation}
Also note the appearance of these averagings in the spectral values of the quantum observable $A_f$:
\begin{equation}\label{outcome1}
\mbox{Sp}(f) = \left\{ \langle f \rangle_{+} \pm \sqrt{\left(\langle f \rangle_{-}\right)^2 + \vert \langle{\pmb \beta } |  {\pmb F} | {\pmb \alpha }\rangle\vert^2}\right\}.
\end{equation}
Just remark that  if vector ${\pmb \alpha } = (1,0, \cdots , 0)$ is part of the canonical basis and ${\pmb \beta } = \left(0, \beta_2, \cdots, \beta_n \right)$
is unit vector orthogonal to ${\pmb \alpha }$, then $A_f$ is diagonal and $\mbox{Sp}(f)$ is trivially reduced to $\left( f(x_1),  \langle  {\pmb F} \rangle_{\pmb \beta }\right)$.
The upper symbols for Pauli matrices read in vector form as
\begin{equation}
 \hat{\pmb \sigma}_0 = \begin{pmatrix}
 1  \\
1\\ 
 1 \\ 
 1
\end{pmatrix}, \ \hat{\pmb \sigma}_1 = {\mathcal C}^{-1}\begin{pmatrix}
 0  \\
0\\ 
 1 \\ 
 0
\end{pmatrix}, \ \hat{\pmb \sigma}_2 = {\mathcal C}^{-1}\begin{pmatrix}
 0  \\
0\\ 
 0 \\ 
 -1
\end{pmatrix}, \ \hat{\pmb \sigma}_3 = {\mathcal C}^{-1}\begin{pmatrix}
 1  \\
-1\\ 
 0 \\ 
 0
\end{pmatrix}.  
\end{equation}
On the other hand, and for any $N$,  components of the lower symbol of $A_f$ are given in terms of another probability distribution in which
the importance of each one is precisely doubled relatively to its counterpart in (\ref{probf}):
\begin{equation}
\label{ }
\langle x_l| A_f |x_l \rangle = \check{A}_f (x_l) =  \sum_{i=1}^N \varpi_{li} f(x_i),
\end{equation}
with
\begin{equation}
\varpi_{ll} = \vert \alpha_l \vert^2 + \vert \beta_l \vert^2, \ \varpi_{li} = \frac{\vert \overline{\alpha_l} \alpha_i +
\overline{\beta_l} \beta_i\vert^2}{\vert \alpha_l \vert^2 + \vert \beta_l \vert^2}, \ i\not= l.
\end{equation}
Note that the matrix $(\varpi_{li} ) $ is  stochastic. 
As a matter of fact, components of lower symbols of Pauli matrices are given by:
\begin{align}
\check{\sigma}_0 (x_l) &= 1, &  \check{\sigma}_1 (x_l) & = \frac{2 \Re\left(\overline{\alpha_l} \beta_l\right)}{\vert \alpha_l
\vert^2 + \vert \beta_l \vert^2}, \\ \check{\sigma}_2 (x_l) & = \frac{2 \Im\left(\overline{\alpha_l} \beta_l\right)}{\vert \alpha_l
\vert^2 + \vert \beta_l \vert^2}, & \check{\sigma}_3 (x_l) &= \frac{\vert \alpha_l \vert^2 - \vert \beta_l \vert^2}{\vert \alpha_l
\vert^2 + \vert \beta_l \vert^2}.
\end{align}
Hidden behind this formal game lies an interpretation resorting to Hermitian geometry probability \cite{}. For instance, consider 
$X= \{x_i \}$ as a set of N real numbers. One then can view the real-valued function $f$ defined by $f(x_i) = x_i$ as the \textit{position} observable,
the measurement of which on the quantum level determined by the choice  of ${\pmb \alpha }= \{\alpha_i \}, {\pmb \beta }= \{\beta_i\}$
has the two possible outcomes given by (\ref{outcome1}). Moreover,  the \textit{position} $x_l$ is privileged to a certain (quantitative) extent in the 
expression of the average value of the \textit{position} operator when computed in state  $ |x_l \rangle$.

Before ending this section, let us examine the lower-dimensional cases $N=2$ and $N=3$. When  $N=2$ the basis change 
(\ref{osN}) reduces to a $U(2)$ transformation with $SU(2)$ parameters $\alpha = \alpha_1$, $\beta = - \overline{\beta}_1$,
$\vert \alpha \vert^2 - \vert \beta \vert^2 = 1$, and some global phase factor. The operator (\ref{aefN})  simplifies as
\begin{equation}
A_f = f_+ \I + 
f_- \begin{pmatrix}
\vert \alpha \vert^2  - \vert \beta \vert^2 & - 2  \alpha \beta\\ 
-2 \overline{\alpha \beta} 
&   \vert \beta \vert^2 - \vert \alpha \vert^2  \end{pmatrix},
\end{equation}
with $f_{\pm} := (f(x_1) \pm f(x_2))/2 $. We now have a two-dimensional commutative  algebra  of ``observables'' $A_f$, 
generated by the identity matrix $\I = \sigma_0$ and the $SU(2)$ transform of $\sigma_3$: $\sigma_3 \rightarrow g \sigma_3
g^{\dagger}$ with 
$g = \bigl( \begin{smallmatrix} \alpha & \beta \\ -\bar{\beta} & \bar{\alpha} \end{smallmatrix}\bigr) \in SU(2)$. As is easily
expected in this case, lower symbols reduce to components:
\begin{equation}
\langle x_l| A_f |x_l \rangle = \check{A}_f (x_l) = f(x_l), \ l= 1,2.
\end{equation}
Finally, it is interesting to consider the $N=3$ case when all considered vector spaces are real. The basis change (\ref{osN})
involves  four real independent parameters, say $\alpha_1, \alpha_2, \beta_1$, and $\beta_2$, all with modulus $< 1$.  The
counterpart of  (\ref{matc}) reads here as
\begin{equation}\label{matcr}
\mathcal C_3 = \begin{pmatrix}
( \alpha_1 )^2 & ( \alpha_2 )^2 & 1 - (\alpha_1 )^2 - ( \alpha_2 )^2  \\
( \beta_1 )^2 & ( \beta_2 )^2 & 1 - (\beta_1 )^2 -( \beta_2 )^2\\ 
  \alpha_1 \beta_1 &   \alpha_2 \beta_2 &  - \alpha_1 \beta_1 - \alpha_2 \beta_2
\end{pmatrix}
\end{equation}
If $\det \mathcal C_3 = ( \alpha_1 \beta_2 - \alpha_2 \beta_1)( \beta_1 \beta_2- \alpha_1 \alpha_2 )\not= 0 $, then one has
uniqueness of upper symbols of Pauli matrices  $\sigma_{1}, \sigma_{3}$, and $\sigma_{0} =Ê\I$ which form a basis of the
three-dimensional Jordan algebra of
real symmetric $2\times 2$-matrices. These upper symbols read in vector form as
\begin{equation}
 \hat{\pmb \sigma}_0 = \begin{pmatrix}
 1  \\
1\\ 
 1  
 \end{pmatrix}, \ \hat{\pmb \sigma}_1 = {\mathcal C_3}^{-1}\begin{pmatrix}
 0  \\
0\\ 
 1  
 \end{pmatrix}, \ \hat{\pmb \sigma}_3 = {\mathcal C_3}^{-1}\begin{pmatrix}
 1  \\
-1\\ 
 0  
\end{pmatrix}.  
\end{equation}

Finally, the extension of this quantization formalism to $N'$-dimensional subspaces of the original $L^2(X, \mu) \simeq \C^N$
appears as being straighforward on a technical if not interpretational level.

\section{Quantum processing of the unit interval}

\subsection{ Quantization with finite subfamilies of Haar wavelets}

 Further simple examples of  quantization 
 are provided when we deal with the
unit interval $X = \lbrack 0,1 \rbrack$ of the real line and its
associated Hilbert space $L^2 \lbrack 0,1 \rbrack$.

Let us  start out by simply selecting the
two first elements of the orthonormal Haar basis \cite{daube}, namely
the characteristic function $\mathbf1 (x)$ of the unit interval and
the Haar wavelet:
\begin{equation}\label{haar}
\phi_1(x) = \mathbf1 (x), \ \phi_2 (x) = \mathbf1 (2x) - \mathbf1
(2x-1). \end{equation}
Then we have,
\begin{equation}
\mathcal N(x) = \sum_{n=1}^2 \vert \phi_n(x) \vert^2 = 2 \ \
\mbox{\it a.e.}. \end{equation}
The corresponding coherent states read as
\begin{equation}\label{uics}
| x \rangle = \frac{1}{\sqrt{2}} \left\lbrack \phi_1(x) ~|1 \rangle +
\phi_2(x) ~|2 \rangle \right\rbrack. \end{equation}
To any integrable function $f(x)$ on the interval there corresponds
the linear operator $A_f$ on $\R^2$ or $\C^2$:
\begin{equation}\label{aefui}
\begin{split}
A_f &= 2\int_0^1 dx\, f(x) | x \rangle \langle x | \\ & =
\left\lbrack \int_0^1 dx\, f(x) \right\rbrack \left\lbrack |1\rangle
\langle 1 | + |2\rangle \langle 2 | \right\rbrack + \left\lbrack
\int_0^1 dx\, f(x)\phi_2(x) \right\rbrack \left\lbrack |1\rangle
\langle 2 | + |2\rangle \langle 1 | \right\rbrack, \end{split}
\end{equation}
or, in matrix form with respect to the orthonormal basis
(\ref{haar}), \begin{equation}
A_f = \begin{pmatrix}
\int_0^1 dx\, f(x) & \int_0^1 dx\, f(x)\phi_2(x) \\ \int_0^1 dx\,
f(x)\phi_2(x) & \int_0^1 dx\, f(x) \end{pmatrix} .
\end{equation}
In particular, with the choice $f=\phi_1 $ we recover the identity
whereas for $f=\phi_2 $, $A_{\phi_2} = \bigl( \begin{smallmatrix} 0 &
1 \\ 1 & 0 \end{smallmatrix} \bigr) = \sigma_1$, the first Pauli
matrix. With the choice $f(x) = x^p, \Re e \,p > -1 $,
\begin{equation}
A_{x^p} = \frac{1}{p+1} \begin{pmatrix}
1 & 2^{-p} - 1\\
2^{-p} - 1 & 1
\end{pmatrix} .
\end{equation}

For an arbitrary coherent state $ | x_0 \rangle,\, x_0 \in \lbrack 0,1
\rbrack $, it is interesting to evaluate the average values (lower symbols) of
$A_{x^p} $.  This gives \begin{equation}
\langle x_0 | A_{x^p} | x_0 \rangle = \left\{ \begin{array}{ll}
\frac{2^{-p} }{p+1} & 0 \leq x_0 \leq \frac{1}{2}, \\
\frac{2 - 2^{-p} }{p+1} & \frac{1}{2} \leq x_0 \leq 1 , \end{array}
\right. \end{equation}
the two possible values being  precisely   the
eigenvalues of the above matrix.
Note   the average values of the ``position'' operator:
$\langle x_0 | A_{x} | x_0 \rangle =1/4$ if $0 \leq x_0 \leq \frac{1}{2}$ and $3/4$ if $\frac{1}{2} \leq
x_0 \leq 1$.

Clearly, like in the $N=2$ case of the previous section,  all operators $A_f$ commute, since
they are linear combinations of the identity matrix and the Pauli
matrix $\sigma_1$.
The procedure is easily generalized to higher dimensions. Let us add to the previous set $\{\phi_1, \phi_2\}$
other elements of the Haar basis, say up to ``scale'' $J$ :
\begin{multline}
\label{ }
\lbrace\phi_1(x), \phi_2 (x), \phi_3 (x) = \sqrt 2 \phi_2(2x), \phi_4(x) = \sqrt 2 \phi_2(2x -1), \\
\cdots, \phi_s (x) =2^{j/2} \phi_2(2x - k), 
 \phi_N (x) =  2^{J/2}\phi_2(2x - 2^J + 1) \rbrace,
\end{multline}
where, at given $j = 1, 2, \cdots, J$, the integer $k$ assumes its values in the range $ 0 \leq k \leq 2^j -1$. The total number 
of elements of this orthonormal system is $N = 2^{J+1}$. The expression of (\ref{factor}) is also given by $\mathcal N(x) =  2^{J+1}$, and this clearly diverges at the  limit $J \to \infty$. Then, it is remarkable if not expected that spectral values as well as  average values of the ``position'' operator are given by
$\langle x_0 | A_{x^p} | x_0 \rangle =(2k+1)/2^{J+1}$ for $ k/2^J \leq x_0 \leq (k+1)/2^J$ where $ 0 \leq k \leq 2^J -1$. Our quantization scheme in the present case achieves a dyadic discretization of the localization in the unit interval.

\subsection{A two-dimensional non-commutative quantization of the unit interval}

Now we choose another orthonormal system, in the form of the two first
elements of the trigonometric Fourier basis,

\begin{equation}\label{trig}
\phi_1(x) = \mathbf1 (x), \ \phi_2 (x) = \sqrt{2}\sin{2\pi x}.
\end{equation}
Then we have,
\begin{equation}
\mathcal N(x) = \sum_{n=1}^2 \vert \phi_n(x) \vert^2 = 1 + 2
\sin^2{2\pi x}, \end{equation}
and corresponding coherent states read as \begin{equation}\label{uics2}
| x \rangle = \frac{1}{\sqrt{1 + 2~ \sin^2{2\pi x}}} \left\lbrack |1
\rangle + \sqrt{2}~\sin{2\pi x} ~|2 \rangle \right\rbrack.
\end{equation}
To any integrable function $f(x)$ on the interval, corresponds
the linear operator $A_f$ on $\R^2$ or $\C^2$ (in its matrix form) ,
\begin{equation}
A_f = \begin{pmatrix}
\int_0^1 dx\, f(x) & \sqrt2\int_0^1 dx\, f(x) \sin{2\pi x} \\
\sqrt2\int_0^1 dx\, f(x) \sin{2\pi x} & 2\int_0^1 dx\, f(x)
\sin^2{2\pi x} \end{pmatrix} .
\end{equation}
Like in the previous case, with the choice $f=\phi_1 $ we recover the
identity whereas for $f=\phi_2 $, $A_{\phi_2} = 
\sigma_1$, the first Pauli matrix.

We now have to deal with a non-commutative Jordan algebra of operators
$A_f$, like in the $N=3$ real case of the previous section. It is generated by the identity matrix 
and the two real Pauli
matrices $\sigma_1$ and $\sigma_3$.

\begin{figure}[tb]
\blankbox{0\columnwidth\includegraphics[angle=0,width=9cm]{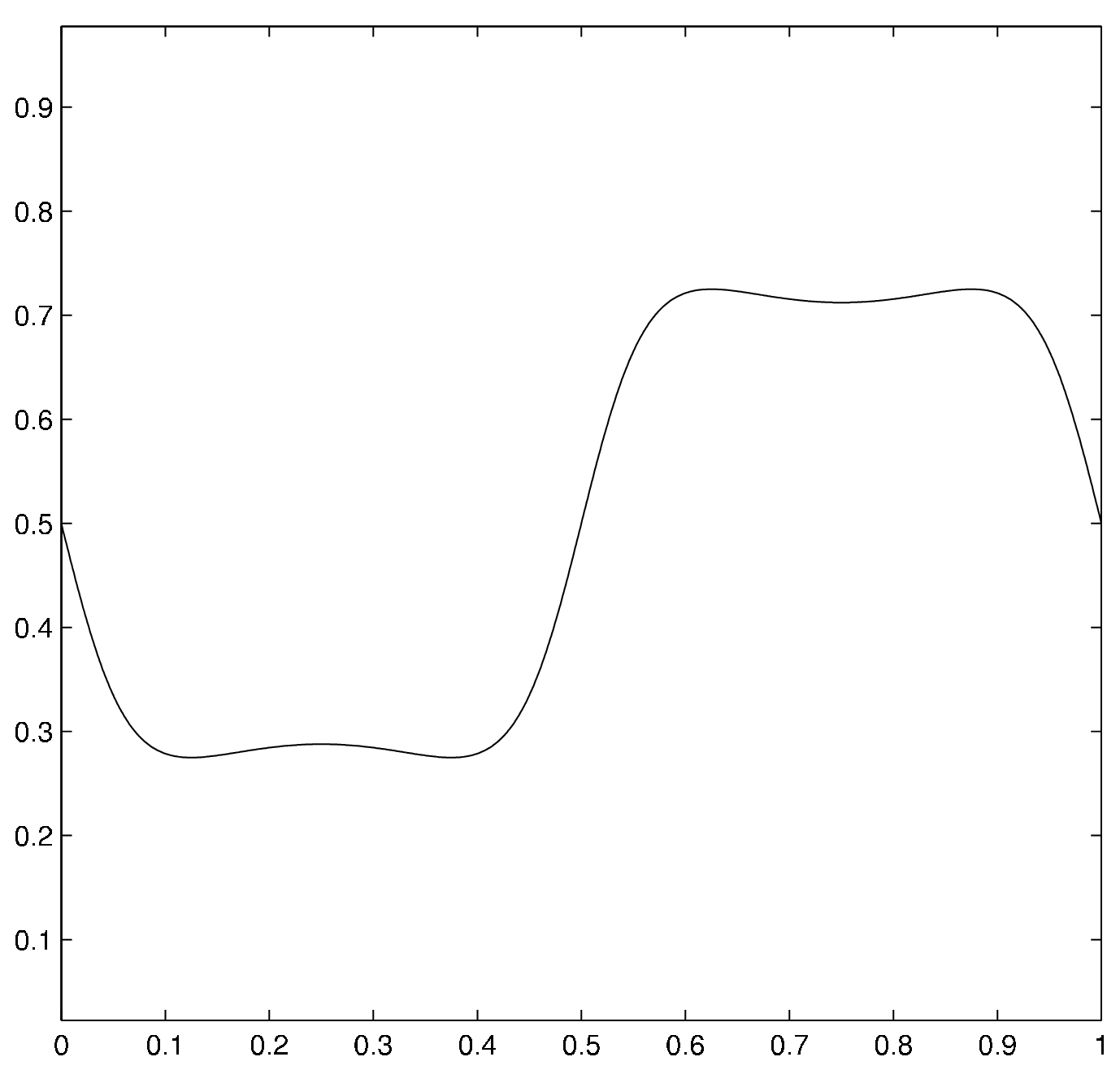}}{20pc}
\caption{Average value $\langle x_0 |A_x | x_0 \rangle$ of \textit{position} operator $A_x$ versus $x_0$,
(compare with eigenvalues of $A_x$).}
\label{ffirstfig}
\end{figure}

In this context, the \textit{position} operator is given by:
\begin{equation*}
A_x = \begin{pmatrix}
\frac{1}{2} & -\frac{1}{\sqrt{2} \pi} \\
- \frac{1}{\sqrt{2} \pi} & \frac{1}{2}\end{pmatrix} ,
\end{equation*}
with eigenvalues $\frac{1}{2}  \pm \frac{1}{\sqrt{2} \pi}$
 Note   its average values  in  function of the coherent state parameter  $x_0 \in \lbrack 0,1 \rbrack $:
\begin{equation*}
\langle x_0 |A_x | x_0 \rangle = \frac{1}{2}  - \frac{2}{\pi} \frac{\sin{2\pi x_0}}{1 + 2~ \sin^2{2\pi x_0}}
\end{equation*}
In Fig.\ref{ffirstfig} we give the curve of $\langle x_0 |A_x | x_0$ in function of $x_0$. It is interesting to compare with the two-dimensional Haar quantization presented in the previous subsection. 

\section{Conclusion}

 The  examples we have given in this contribution are mainly of pedagogical nature. Other examples, specially devoted to Euclidean and pseudo-euclidean spheres will be presented elsewhere, having in view possible connections with  objects of noncommutative geometry (like fuzzy spheres, see for instance \cite{frkr}). They show the extreme freedom we have in analyzing a set $X$ of \textit{data} or \textit{possibilities}  just equipped with a measure by following a quantumlike procedure. The crucial step lies in the choice  of a countable orthonormal subset in $ L^2(X, \mu)$ obeying (\ref{factor}). A $\C^N$ (or $l^2$ if $N = \infty$) unitary transform of this original subset would actually lead to the same specific quantization, and 
 the latter could as well be obtained by using unitarily equivalent \textit{continuous} orthonormal distributions defined within the framework of some Gel'fand triplet. Of course, further structure like symplectic manifold combined with spectral constraints imposed to some specific observables will considerably restrict that freedom and will lead hopefully to a unique solution, like Weyl quantization, deformation quantization, or geometric quantization are able to achieve in specific situations. Nevertheless,  we believe that the generalization of Berezin quantization which has been  described here, and which goes far beyond the context of Classical and Quantum Mechanics, not only will  shed light on the specific nature of the latter, but also will   help to solve  in a simpler way some quantization problems.

\bibliographystyle{amsalpha}

\end{document}